\iffalse Before sending further, pass this check list:
  [ ] find and resolve all "??" and "\ifnote"
  [ ] Spell-check
  [ ] order of using/defining abbreviations
  [ ] citation order (use RefTest)
  [ ] figure order & reference order
  [ ] Check LaTeX  output files (*.log) for warnings
  [ ] Check BibTeX output (screen) for warnings
  [ ] update PACS
  [ ] decide on color figures
  [ ] Re-read paper in the morning

After completing this list, if you made at least one correction,
re-do this check-list from the beginning, until no corrections will
be done.\fi

\documentclass[amsmath,amssymb,prb,preprintnumbers,showpacs,superscriptaddress,twocolumn]{revtex4}

\usepackage{graphicx}
\usepackage{dcolumn}          
\usepackage{bm}               
\usepackage[usenames]{color}  
\usepackage{ulem}             
\usepackage[colorlinks=true, linkcolor=black, citecolor=black, urlcolor=black]{hyperref} 
\usepackage{cancel}	      

\newcommand{\ncco}{Nd$_{2-x}$Ce$_{x}$CuO$_{4-\delta}$ }
\newcommand{\lcco}{La$_{2-x}$Ce$_{x}$CuO$_{4-\delta}$ }

\newif\ifcom
\newif\ifdel

\begin{document}


\title{Importance of grain boundary Josephson junctions in the\\
electron-doped infinite-layer cuprate superconductor Sr$_{1-x}$La$_{x}$CuO$_{2}$}

\author{J.~Tomaschko}
\affiliation{%
  Physikalisches Institut -- Experimentalphysik II and Center for Collective Quantum Phenomena in LISA$^+$,
  Universit\"{a}t T\"{u}bingen,
  Auf der Morgenstelle 14,
  72076 T\"{u}bingen, Germany
}
\author{V.~Leca}
\affiliation{%
  Physikalisches Institut -- Experimentalphysik II and Center for Collective Quantum Phenomena in LISA$^+$,
  Universit\"{a}t T\"{u}bingen,
  Auf der Morgenstelle 14,
  72076 T\"{u}bingen, Germany
}
\affiliation{%
  Faculty of Applied Chemistry and Materials Science,
  University Politehnica of Bucharest,
  Gheorghe Polizu Street 1-7,
  Bucharest 011061, Romania
}
\author{T.~Selistrovski}
\affiliation{%
  Physikalisches Institut -- Experimentalphysik II and Center for Collective Quantum Phenomena in LISA$^+$,
  Universit\"{a}t T\"{u}bingen,
  Auf der Morgenstelle 14,
  72076 T\"{u}bingen, Germany
}
\author{R.~Kleiner}
\author{D.~Koelle}
\email{koelle@uni-tuebingen.de}
\affiliation{%
  Physikalisches Institut -- Experimentalphysik II and Center for Collective Quantum Phenomena in LISA$^+$,
  Universit\"{a}t T\"{u}bingen,
  Auf der Morgenstelle 14,
  72076 T\"{u}bingen, Germany
}
%
\date{\today}

\begin{abstract}
Grain boundary bicrystal Josephson junctions of the electron-doped infinite-layer superconductor Sr$_{1-x}$La$_x$CuO$_2$ ($x = 0.15$) were grown by pulsed laser deposition.
BaTiO$_3$-buffered $24\,^\circ$ [001]-tilt symmetric SrTiO$_3$ bicrystals were used as substrates.
We examined both Cooper pair (CP) and quasiparticle (QP) tunneling by electric transport measurements at temperatures down to 4.2\,K.
CP tunneling revealed an extraordinary high critical current density for electron-doped cuprates of $j_c > 10^3\,$A/cm$^2$ at 4.2\,K.
Thermally activated phase slippage was observed as dissipative mechanism close to the transition temperature.
Out-of-plane magnetic fields $H$ revealed a remarkably regular Fraunhofer-like $j_c(H)$ pattern as well as Fiske and flux flow resonances, both yielding a Swihart velocity of $3.1\cdot10^6\,$m/s.
Furthermore, we examined the superconducting gap by means of QP tunneling spectroscopy.
The gap was found to be V-shaped with an extrapolated zero temperature energy gap $\Delta_0 \approx 2.4\,$meV.
No zero bias conductance peak was observed.
\end{abstract}

\pacs{
61.72.Mm,		
74.25.Sv,		
74.50.+r,   
74.72.Ek    
}

\maketitle

\section{Introduction}
\label{sec:introduction}

Only two families of electron-doped high transition temperature (high-$T_c$) cuprate superconductors are known, the $T'$-compounds\cite{Tokura89, Takagi89} and the infinite-layer (IL) compounds\cite{Siegrist88, Smith91, Er92} $A^\mathrm{II}_{1-x}Ln^\mathrm{III}_x$CuO$_2$ (where $A^\mathrm{II}$ is a divalent alkaline earth metal and $Ln^\mathrm{III}$ is a trivalent lanthanide).
The IL compounds exhibit the simplest crystal structure among all cuprate superconductors and are therefore predestined to examine the nature of high-$T_c$ superconductivity.
However, as the fabrication of high-quality superconducting IL samples is challenging, these compounds have been less examined than any other cuprate superconductor.
High pressure preparation is necessary to stabilize the superconducting IL phase, yielding polycrystalline bulk material\cite{Er91, Ikeda93, Jorgensen93}.
To obtain single crystalline samples, IL thin films were deposited epitaxially on single crystalline templates\cite{Niu92,Adachi92,Karimoto01}.
After it had been shown that compressive strain hinders electron-doping of the CuO$_2$ planes, buffer layers were introduced\cite{Adachi92}, enhancing the $T_c$ close to the maximum of 43\,K\cite{Er92,Karimoto01}.

The importance of grain boundary Josephson junctions (GBJs) in the high-$T_c$ cuprates has been demonstrated by various experiments (see e.g.~reviews\cite{Gross94a,Mannhart01,Hilgenkamp02} and references therein).
Chaudhari \textit{et al.}\cite{Chaudhari88} realized the first single YBa$_2$Cu$_3$O$_{7-\delta}$ (YBCO) GBJs.
They showed that such GBJs act as weak links, which later on has been used e.g.~for the realization of sensitive high-$T_c$ SQUIDs\cite{Koelle99}.
In particular, cuprate GBJs played a decisive role in experimental tests on the determination of the superconducting order parameter symmetry of the cuprates\cite{Tsuei00}.
For example, Alff \textit{et al.}\cite{Alff98} observed an increased conductance across various high-$T_c$ GBJs at low voltage, a so-called zero bias conductance peak (ZBCP), which was explained by the formation of zero energy surface states (so-called Andreev bound states, ABS) due to $d$-wave pairing of the underlying material\cite{Hu94,Kashiwaya00, Deutscher05}.
Tsuei \textit{et al.}\cite{Tsuei94} revealed $d$-wave pairing in YBCO with superconducting rings containing three GBJs.
Schulz \textit{et al.}\cite{Schulz00} realized an all high-$T_c$ 0-$\pi$-SQUID comprising two YBCO GBJs with phase shifts of 0 and $\pi$, respectively, to prove $d$-wave pairing in YBCO, and later on a similar experiment was done for the electron-doped $T'$-compound La$_{2-x}$Ce$_x$CuO$_{4-\delta}$\cite{Chesca03}.
And Lombardi \textit{et al.}\cite{Lombardi02a} found an oscillatory dependence of the critical current density $j_c$ on GBJ misorientation angle, as an indication of $d$-wave paring in YBCO.

However, so far all experiments on IL cuprates have only been performed on bulk polycrystals or single crystalline thin films, but no Josephson devices have been fabricated.
Very recently, we reported on thin film planar Sr$_{1-x}$La$_{x}$CuO$_2$ (SLCO)/Au/Nb junctions\cite{Tomaschko11}, which showed quasiparticle (QP) tunneling,
but no Cooper pair (CP) tunneling.
%
In this work, we report on the fabrication and characterization of SLCO bicrystal GBJs.
We found high $j_c$ well below $T_c$, thermally activated phase slippage close to $T_c$, very regular Fraunhofer-like $j_c$ vs magnetic field $H$ patterns as well as Fiske and flux flow resonances as remarkable features of CP tunneling.
QP tunneling essentially revealed a well-defined, V-shaped gap but did not show any ZBCP.
Such devices may give new insights into basic properties of the IL cuprates and allow for a comparison with the properties of GBJs based on other hole- and electron-doped cuprates.
Moreover, the pairing symmetry of the IL cuprates is still an open question and phase-sensitive experiments based on SLCO GBJs could help finding the answer.

\section{Sample fabrication and experimental setup}
\label{sec:fabrication}

To fabricate SLCO GBJs, we deposited SLCO thin films on BaTiO$_3$-buffered symmetric [001]-tilt SrTiO$_3$ bicrystals with misorientation angle $\theta=24\,^\circ$ by pulsed laser deposition, cf.~Ref.~[\onlinecite{Tomaschko11}].
Further details will be described elsewhere\cite{Tomaschko11b}.
Four chips, each with seven junctions, have been fabricated and characterized.
All SLCO GBJs showed comparable properties, which verifies reproducibility of our data.
For simplicity, data presented in this work stems from one chip.
The SLCO films with thickness $t\sim 25\,$nm are $c$-axis oriented, as confirmed by x-ray diffraction, i.e., current flow is restricted to the $ab$-plane.
Each chip was patterned via standard photolithography and argon ion milling to create junctions with widths $w$ ranging between 10 and $1000\,\mu$m.
Electric transport measurements were performed in a 4-point configuration, with the sample mounted inside a noise filtered, magnetically and radio frequency shielded probe in a liquid helium dewar.

\section{Electric Transport properties at zero magnetic field}
\label{sec:zero-field}

\begin{figure}[htb!] 
\centering
\includegraphics[width=1.00\columnwidth]{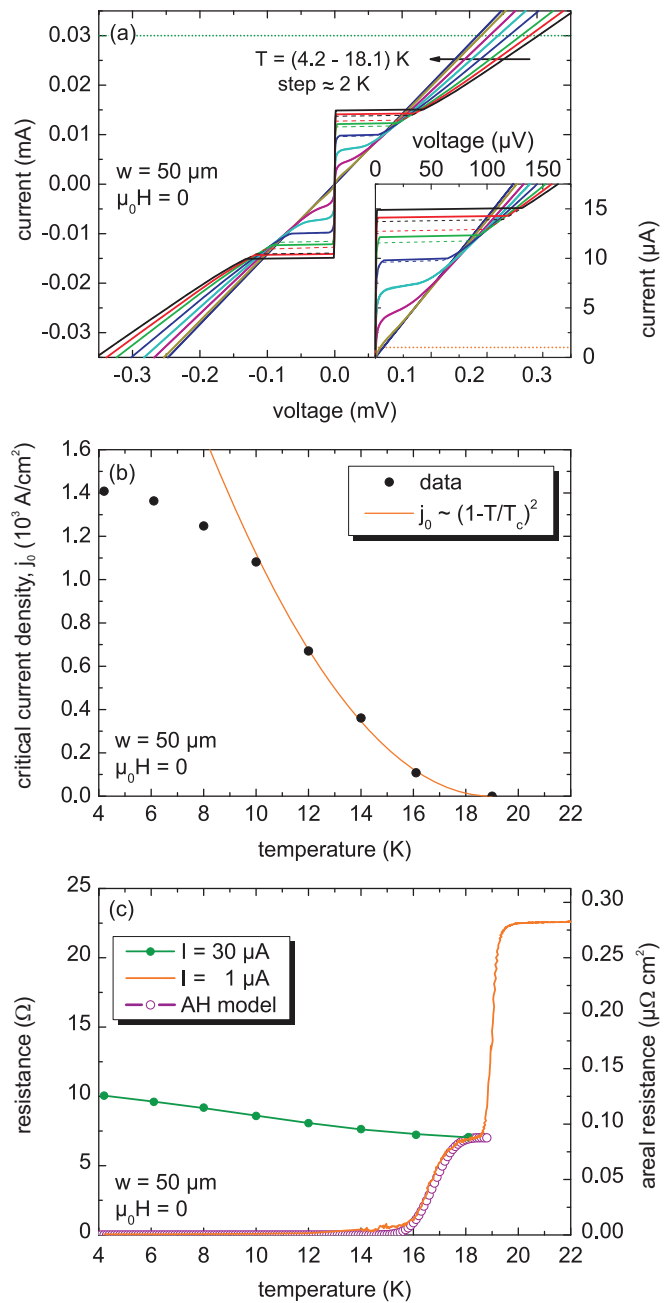}
\caption{(Color online)
Electric transport characteristics at $H=0$ for 50\,$\mu$m wide SLCO GBJ.
(a) Current-voltage curves at different temperatures.
Inset shows an expanded view.
Curves for decreasing $I$ are shown as dashed lines.
The horizontal dotted lines indicate the values used for $I$ in (c).
(b) Critical current density $j_0$ vs $T$.
Values (full symbols) were determined by RCSJ fits of the IVCs.
The line is a parabolic fit of the data for $T \ge 10\,$K.
(c) Resistance vs temperature, measured with different bias currents.
The curve shown for $I=30\,\mu\mathrm{A}\approx2I_c(4.2\,\mathrm{K})$ corresponds to the normal resistance $R_n$ of the junction.
Open symbols show the calculated resistance according to the Ambegaokar-Halperin (AH) model.
}
\label{fig:fig01}
\end{figure}

Figure \ref{fig:fig01}(a) shows bias current $I$ vs voltage $V$ characteristics (IVCs) of a 50\,$\mu$m wide junction at various temperatures $T$ and $H$=0.
The IVCs are resistively and capacitively shunted junction (RCSJ)-like and do not show a significant excess current.
At low $T$ there is a small hysteresis [cf.~inset of Fig.~\ref{fig:fig01}(a)], corresponding to a Stewart-McCumber parameter $\beta_c = 2\pi I_0 R_n^2 C/\Phi_0$ slightly above 1.
Here, $I_0$ is the maximum Josephson current in the absence of thermal noise, and $R_n$ and $C$ are the normal resistance and capacitance of the junction, respectively; $\Phi_0$ is the magnetic flux quantum.
By comparison of measured IVCs with numerical RCSJ simulations including thermal noise, we determined $\beta_c(T)$ and the noise parameter (thermal energy divided by Josephson coupling energy) $\Gamma(T)\equiv 2\pi k_B T / (I_0(T) \Phi_0)$; $k_B$ is the Boltzmann constant.

With the 4.2\,K values $\beta_c=3.5$ and $\Gamma=0.01$ and the measured $R_n = 10\,\Omega$, we find $I_0=17.6\,\mu$A and $C=0.65\,$pF.
The 4.2\,K value for $I_0$ corresponds to a critical current density $j_0 \approx 1.4\,$kA/cm$^2$.
This value is two orders of magnitude below $j_0$ values for corresponding ($\theta=24\,^\circ$) YBCO GBJs\cite{Hilgenkamp98}, but two orders of magnitude above $j_0$ of GBJs from electron-doped $T'$-compounds such as \ncco (NCCO)\cite{Kleefisch98} and \lcco (LCCO)\cite{Wagenknecht08a}, thus, probably the highest reported $j_0$ value for electron-doped cuprate GBJs.
The 4.2\,K value for $C$ corresponds to a capacitance per junction area $C/A \approx 50\,\mu$F/cm$^2$, which is by a factor of $\sim 20$  larger than typical $C/A$ values of YBCO GBJs\cite{Winkler94,Hilgenkamp02}.

With increasing $T$, $j_0$ decreases non-linearly as shown in Fig.~\ref{fig:fig01}(b).
For 10\,K\,$\le$\,$T$\,$\le$\,19\,K, we find a power law behavior $j_0 \propto (1 - T/T_c)^\alpha$ with $\alpha = 2.0 \pm 0.1$.
Such quadratic behavior was explained by a reduction of the order parameter $\Delta$ at the GB interface due to the small coherence length $\xi$ of the cuprate superconductors\cite{Deutscher87}.
For lower $T$, high-$T_c$ GBJs usually exhibit a linear $j_0(T)$ dependence\cite{Hilgenkamp02}.
Contrary to that, we observed a monotonous decrase in $|dj_0/dT|$ with decreasing $T$ below $\sim 9\,$K.
This resembles the behavior of an early YBCO GBJ, reported by Mannhart \textit{et al.}\cite{Mannhart88}, in good agreement with the Ambegaokar-Baratoff relation\cite{Ambegaokar63}, which is valid for superconductor-insulator-superconductor (S/I/S) junctions.

Figure~\ref{fig:fig01}(c) shows $R(T)$ curves measured with different bias currents $I$.
To access the normal resistance $R_n$ of the GBJ, in one case an overcritical current $I = 30\,\mu$A was applied.
We found, that $R_n$ increases from 7.1 to 10.1$\,\Omega$ when $T$ is lowered from 19 to 4.2\,K, corresponding to an areal resistance $\rho_n \equiv R_n A = 0.09$ -- $0.13\,\mu\Omega\,\mathrm{cm}^2$.
Thus, $\rho_n$ is inbetween the values reported for YBCO GBJs ($10^{-3}$ -- $10^{-2}\,\mu\Omega$cm$^2$)\cite{Hilgenkamp98,Ransley04} and the values for GBJs from the $T'$ compounds NCCO (1 -- $10\,\mu\Omega$cm$^2$)\cite{Kleefisch98} and LCCO
(30 -- $130\,\mu\Omega$cm$^2$) GBJs\cite{Wagenknecht08a}.

We further find $I_0 R_n$  $\sim$ 0.18\,mV at 4.2\,K, which is comparable to values for NCCO GBJs\cite{Kleefisch98}, but 1\,-\,2 orders of magnitude smaller than the products ($\sim 1\,-\,8\,$mV) reported for LCCO GBJs\cite{Wagenknecht08a} or YBCO GBJs\cite{Hilgenkamp98}.
We note that our values of $I_0R_n$ and $j_0$ fall right onto the scaling line $I_0 R_n \propto 1/\rho_n^{1.5}$ shown by Gross and Mayer\cite{Gross91a} for YBCO GBJs, which was suggested to be related to the oxygen stoichiometry at the barrier.
Thus, we speculate that our small $I_0 R_n$ product is a fingerprint of oxygen vacancies at the barrier due to vacuum annealing\cite{Tomaschko11,Tomaschko11b} of the as-deposited SLCO films, which would also be in line with the conclusions of Hilgenkamp and Mannhart\cite{Hilgenkamp02}.
The intrinsically shunted junction (ISJ) model\cite{Gross91a,Hilgenkamp02} is based upon such barrier defects, which are supposed to form localized states.
On the one hand, localized states suppress $\Delta$ at the grain boundary interface.
On the other hand, they enable resonant QP tunneling across the barrier.
The former point is in accordance with our observed quadratic $j_0(T)$ behavior close to $T_c$ and it further explains the small $I_0 R_n$ products.
The latter one explains the small $\rho_n$, because resonant tunneling is consistent with a highly transparent barrier.
Finally, we want to remark, that our small $I_0 R_n$ product is comparable to values reported for those $T'$-compound GBJs\cite{Kleefisch98,Wagenknecht08a} which did \textit{not} exhibit a ZBCP.

We next discuss the resistive transition, which is also shown in Fig.~\ref{fig:fig01}(c) as the $R(T)$ curve measured with small bias current $I=1\,\mu$A.
Upon decreasing $T$, we first find a sharp decrease in $R$, which we associate with the transition of the SLCO film to the superconducting state, with $T_c = 19.0\,$K and width of the resistive transition $\Delta T_c \approx 0.5\,$K.
We note that the observed value of $T_c$ is only about half of the maximum $T_c$ value reported for SLCO \cite{Er92,Karimoto01}.
This can be attributed to non-optimum doping of our SLCO films ($x = 0.15$) and/or to strain effects due to the lattice mismatch of the BaTiO$_3$ buffer layer and SLCO ($a_\mathrm{BTO} = 3.997\,$\AA, $a_\mathrm{SLCO}^\mathrm{bulk} = 3.949\,$\AA)\cite{Karimoto04,Tomaschko11}.
For $T<T_c$ down to $\sim 15\,$K, we observe a ''foot structure`` in $R(T)$.
Such behavior has been observed e.g.,in YBCO and NCCO GBJs before and was attributed to thermally activated phase slippage (TAPS)\cite{Gross90, Kleefisch98}, as described by Ambegaokar and Halperin (AH)\cite{Ambegaokar69}.
Within the AH model, a finite resistance due to TAPS is given by $R_p(T) = R_n(T)\cdot J_0^{-2}[\Gamma^{-1}(T)]$.
Here, $J_0[x]$ is the modified Bessel function of the first kind.
We calculate $R_p(T)$ using the quadratic $j_0(T)$ dependence close to $T_c$, cf.~Fig.~\ref{fig:fig01}(b), and the measured $R_n(T)$.
Similar calculations for other SLCO GBJs with different widths also yielded very good agreement.
Thus, we verified that TAPS is present in our samples and responsible for the finite slope of the $I(V)$ curves ($dI/dV|_{V=0} \neq \infty$) close to $T_c$ [cf.~Fig.~\ref{fig:fig01}(a)].

\section{Electric Transport properties vs magnetic field}
\label{sec:field-dependence}

\begin{figure}[htb!] 
\centering
\includegraphics[width=1.00\columnwidth]{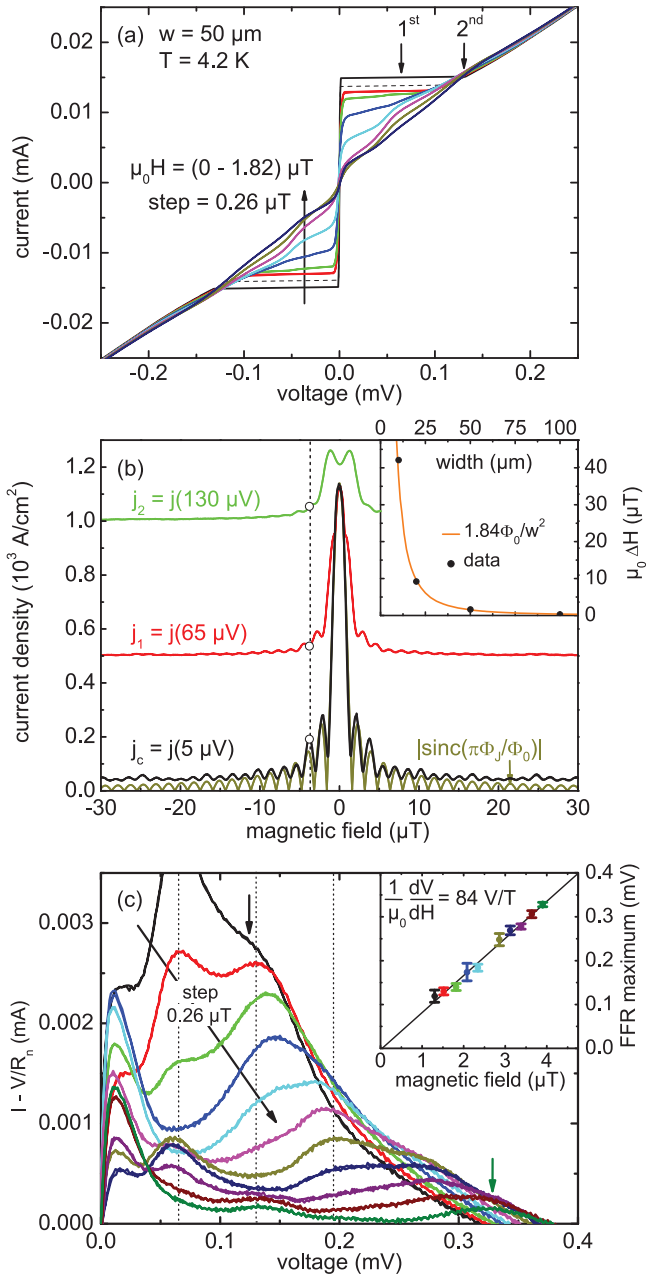}
\caption{(Color online)
(a) Current-voltage curves of 50\,$\mu$m wide SLCO GBJ for different values of $H \parallel c$-axis ($T = 4.2\,$K).
(b) Magnetic field dependence of the critical current density $j_c$ and the currents $j_1$ and $j_2$ at the $1^\mathrm{st}$ and $2^\mathrm{nd}$ Fiske resonance, respectively.
For comparison with $j_c(H)$, an ideal Fraunhofer patten is also shown.
The inset shows the oscillation period $\mu_0\Delta H$ of $j_c(H)$ vs width of four SLCO GBJs as full circles.
The theoretical dependence\cite{Rosenthal91} $\mu_0\Delta H = 1.84 \Phi_0 / w^2$ is shown as solid line.
(c) Current-voltage curves, with the normal current $I_n = V/R_n$ subtracted from the bias current, for $\mu_0 H = 1.3$ - $3.9\,\mu$T.
Vertical lines indicate Fiske resonances and arrows indicate the position of the flux flow resonance (FFR) maximum (for 1.3 and 3.9\,$\mu$T).
Inset shows the linear magnetic field dependence of the FFR maximum.
}
\label{fig:fig02}
\end{figure}

Figure \ref{fig:fig02}(a) shows IVCs of the $50\,\mu$m wide junction at different magnetic fields $H$ and constant $T = 4.2\,$K.
$H$ was applied perpendicular to the film plane.
The modulation of the critical current density $j_c(H)$ (measured with a voltage criterion $V_c=5\,\mu$V) is shown in Fig.~\ref{fig:fig02}(b) (lower curve).
We observed a remarkably regular Fraunhofer-like pattern with oscillations visible throughout the entire scanned field range ($\pm 65\,\mu$T) (here, we only show the pattern for $\left| \mu_0 H \right| \leq 30\,\mu$T for clarity).
The oscillation period $\mu_0\Delta H$ for this junction is $1.6\,\mu$T.
In the thin film limit ($\lambda_L^2 / t \gg w$; $\lambda_L$ is the London penetration depth), a perpendicular field is focused into the GBJ by the superconducting electrodes and the modulation period becomes \cite{Rosenthal91} $\mu_0\Delta H = 1.84 \Phi_0 / w^2$, which is \textit{independent of} $\lambda_L$.
By analyzing four junctions with different widths $w$ on the same chip, we verified this dependence, as shown in the inset of Fig.~\ref{fig:fig02}(b).

At $\left| V \right| \approx 65$ and 130$\,\mu$V, resonant features appear in the IVCs (cf.~arrows in Fig.~\ref{fig:fig02}(a), labeled as 1$^\mathrm{st}$ and 2$^\mathrm{nd}$).
The amplitude of enhanced current (as compared to a linear IVC) shows an oscillatory dependence on $H$ with the same period as  $j_c(H)$.
This is shown in Fig.~\ref{fig:fig02}(b), where we also plot the current densities $j_1\equiv j(65\,\mu\mathrm{V})$ and $j_2\equiv j(130\,\mu\mathrm{V})$ at the $1^\mathrm{st}$ and $2^\mathrm{nd}$ resonance, respectively.
The first resonance has its minima when $j_c(H)$ and the second resonance are near their maxima [cf.~vertical dotted line in Fig.~\ref{fig:fig02}(b)], characteristic for Fiske resonances\cite{Coon65}, i.e.~standing electromagnetic waves in the junction.
The Fiske resonances appear at equidistant voltages $V_n = \Phi_0 c_S n /2w$, where $c_S$ denotes the Swihart velocity\cite{Swihart61} and $n = 1,2,3\ldots$.
From the measured $V_n$ we find $c_S = 3.1 \cdot 10^6\,$m/s, which is comparable to $c_S^\mathrm{YBCO} \approx 10^7\,$m/s found for YBCO GBJs\cite{Winkler94,Zhang95}.
For the 10, 20, and $100\,\mu$m wide junctions the $1^\mathrm{st}$ Fiske resonance appeared at $V_1 = 316$, 167, and $31\,\mu$V, respectively, yielding $c_S = 3.1$, 3.2, and $3.0 \cdot 10^6\,$m/s.

In Fig.~\ref{fig:fig02}(c), the current $I_n = V/R_n$ has been subtracted from the bias current $I$ to show the resonances in the IVCs (for various values of $H$) more clearly.
For some field values, the positions of the Fiske resonances are marked by vertical dashed lines.
Besides Fiske resonances, a resonance is visible with its maximum position $V_m$ shifting $\propto H$ (cf.~arrows and inset).
The peak height of this resonance decreases monotonically with $H$, which is indicative of a Josephson flux-flow resonance (FFR)\cite{Eck64}.
The peak position of the FFR in \textit{thin film} GBJs is given by\cite{Zhang95} $V_m = d_B\,c_S\,\mu_0H\,F$, where $d_B$ denotes the effective barrier thickness and
$F \approx \Phi_0 / (w\,d_B\,\mu_0\Delta H)$ accounts for flux focusing\cite{Rosenthal91}.
The inset in Fig.~\ref{fig:fig02}(c) shows $V_m$ vs $\mu_0H$.
From the slope  $V_m / \mu_0H = (84 \pm 2)\,$V/T and $\mu_0\Delta H = 1.6\,\mu$T we extract $c_S = (3.2 \pm 0.1) \cdot 10^6\,$m/s, in agreement with the value determined from Fiske resonances.

\section{Quasiparticle tunneling spectra}
\label{sec:QP-tunneling}

\begin{figure}[tb] 
\centering
\includegraphics[width=1.00\columnwidth]{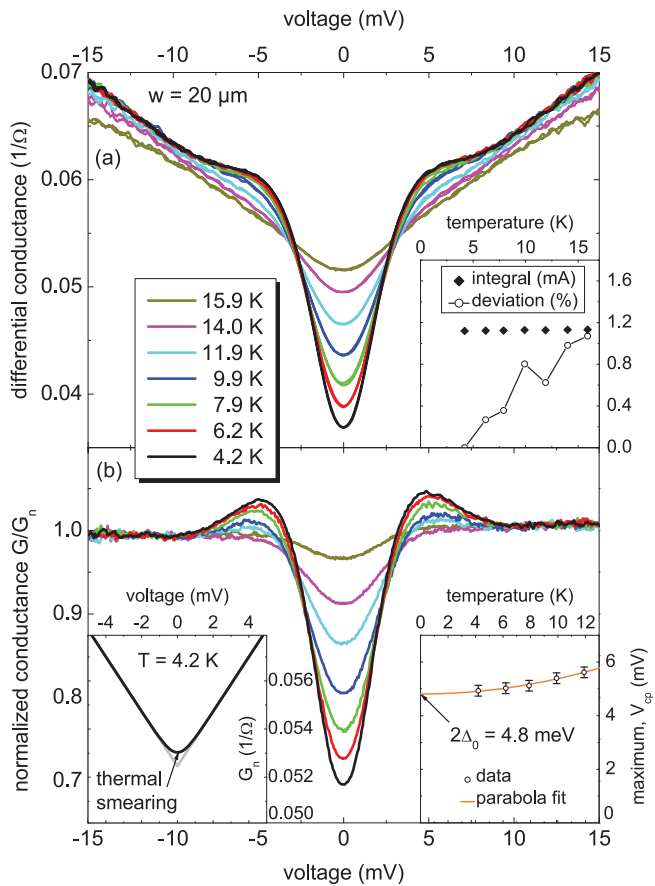}
\caption{(Color online)
(a) Differential conductance $G(V)$ of $20\,\mu$m wide SLCO GBJ for different values of $T$.
The inset shows the integral of the $G(V)$ curves vs $T$ for $-15\,\mathrm{mV} \leq V \leq 15\,\mathrm{mV}$ and the deviation from the integral at 4.2\,K.
(b) Normalized conductance $G/G_n(V)$ at various temperatures.
Left inset shows the normal state conductance $G_n(V)$ at 4.2\,K used for normalization.
The development of the coherence peak position $V_{cp}$ with temperature is illustrated in the right inset.
}
\label{fig:fig03}
\end{figure}

We now turn to quasiparticle tunneling spectra.
Figure \ref{fig:fig03}(a) shows differential conductance curves, $G(V) = dI/dV(V)$, of the $20\,\mu$m wide junction for voltages $\left| V \right| \leq 15\,$mV.
The curves were measured with lock-in technique.
Coherence peaks are clearly visible at voltages near $\pm$5\,mV.
To show that  $G(V)$ indeed probes the density of states (DOS) we have integrated the $G(V)$ curves between $-15\,\mathrm{mV} < V < 15\,\mathrm{mV}$, cf.~inset of Fig.~\ref{fig:fig03}(a).
The maximum deviation from the value at 4.2\,K is $\sim 1\,$\%, i.e., within experimental accuracy, the total DOS is constant.
At $|V|>10$mV $G(V)$ increases linearly with $V$.
Such a V-shaped background conductance $G_n(V)$ has also been reported for $T'$-compound GBJs\cite{Kleefisch01,Wagenknecht08} and is indicative of a normal state DOS increasing linearly with energy.
To determine $G_n(V,T)$ we use the expression $G_n(V,T) = e^{-1} \frac{d}{dV} \int^{\infty}_{-\infty} {N_n(E) [f(E-eV)-f(E)]} dE$, with $N_n(E) \propto a \cdot \left| E \right| + b$ and $f(E) = [1 + \exp(E/k_B T)]^{-1}$, which is fitted to the measured $G(V)$ curves (for $|V|>10\,$mV) for different values of $T$.
The resulting $G_n(V,T)$ exhibit some rounding at low voltages, as illustrated in the left inset of Fig.~\ref{fig:fig03}(b) for $T = 4.2\,$K.
Figure \ref{fig:fig03}(b) shows the normalized conductance $G(V)/G_n(V)$ for temperatures between 4.2\,K and 15.9\,K.

For a fully gapped superconductor the subgap conductivity is U-shaped while for an order parameter with nodes a V-shape is obtained\cite{Won94}.
Thermal smearing\cite{Noce96} and lifetime limiting processes\cite{Dynes78} are rounding the spectra but do not change the substantial shape of the subgap conductivity.
Both Fig.~\ref{fig:fig03}(a) and (b) show that in the subgap regime the conductance is V-shaped.
Moreover, the V-shape becomes more pronounced for decreasing temperature.
These findings suggest that nodes are present in the superconducting order parameter.

The normalized $G(V)/G_n(V)$ curves revealed a slight increase of the coherence peak position $V_{cp}$ with increasing $T$ [cf.~right inset in Fig.~\ref{fig:fig03}(b)].
Such behavior is usually explained by thermal smearing, although, according to BCS theory, the increase of $V_{cp}$ should be more pronounced\cite{Noce96}.
$T'$-compound GBJs, however, even revealed a \textit{decrease} of $V_{cp}$ with increasing $T$\cite{Kleefisch01,Wagenknecht08}.
Furthermore, in GBJs $V_{cp}(0) = 2 \Delta_0/e$\cite{Won94, Noce96}.
A parabolic fit extrapolated to $V_{cp}(0) = (4.8 \pm 0.2)$\,mV, yielding $\Delta_0 = (2.4 \pm 0.1)\,$meV for the $20\,\mu$m wide junction.
Other junctions with different widths revealed essentially the same value.
We thus find a reduced gap ratio $2 \Delta_0 / (k_B T_c) = 3.0 \pm 0.2$, which is somewhat lower than the BCS value 3.5\cite{Bardeen57}.
%
%
We note that our data are in contrast to low temperature scanning tunneling microscopy (STM) results reported for polycrystalline Sr$_{0.9}$La$_{0.1}$CuO$_2$ bulk samples\cite{Chen02} with $T_c = 43\,$K, where the gap was determined to $\Delta_0 = (13 \pm 1)\,$meV, yielding $2 \Delta_0 / (k_B T_c) = 7.0 \pm 0.5$.
However, our data are in line with results obtained on other $T'$-compound thin film GBJs, where $2 \Delta_0 / (k_B T_c) = 2.8$\,-\,3.5\cite{Kleefisch01,Wagenknecht08}.
Furthermore, according to an empirical dependence of the reduced gap ratio on $T_c$, a ratio of $\sim 4$ is expected for samples with $T_c = 19\,$K (cf.~Wei \textit{et al.}\cite{Wei98} and the literature cited therein), which is also close to our data.

The conductance spectra of our junctions did not show a zero bias conductance peak (ZBCP)\cite{Hu94, Kashiwaya95, Tanaka95}.
For Nd$_{2-x}$Ce$_x$CuO$_{4-\delta}$, the absence of a ZBCP in GBJ QP spectra has primarily been interpreted as evidence for $s$-wave pairing\cite{Alff98}.
However, subsequent experiments identified \ncco as $d$-wave superconductor\cite{Ariando05} indicating that the ZBCP has been suppresed, e.g.~by strong disorder at the barrier, reducing the QP mean free path $l_0$ to a value below the in-plane coherence length $\xi_\mathrm{ab}$ and therefore suppressing the constructive interference of electron- and hole-like QPs forming Andreev bound states\cite{Aprili98}.
We also refer to the work of Giubileo \textit{et al.}\cite{Giubileo10} who performed point contact spectroscopy on Pr$_{1-x}$LaCe$_x$CuO$_{4-y}$ crystals.
They showed, that depending on the barrier strength $Z$, different conductance regimes were accessible, the high-$Z$ tunneling regime and the low-$Z$ contact regime, where ZBCPs only occured in the latter one.
Thus, from the absence of a ZBCP in our QP spectra, we cannot conclude $s$-wave pairing.

\section{Conclusions}
\label{sec:conclusions}

In summary, we have fabricated high quality grain boundary Josephson junctions (GBJs) from the electron-doped infinite-layer superconductor Sr$_{1-x}$La$_x$CuO$_2$ ($x = 0.15$) deposited on $24\,^\circ$ symmetric [001]-tilt SrTiO$_3$ bicrystals.
While in many respects these junctions are comparable to GBJs made of other cuprates there are also differences.
For example, the Josephson critical current density of up to $1.4\cdot10^3\,$A/cm$^2$ at 4.2\,K is remarkably high for electron doped cuprates.
The magnetic field dependence of the critical current follows a nearly perfect Fraunhofer pattern which is quite unusual for $24\,^\circ$ cuprate grain boundary junctions.
As for other cuprates, the quasiparticle spectra of our GBJs are V-shaped in the subgap regime indicative of a superconducting order parameter with nodes.
For a $d$-wave order parameter we would have expected zero bias conductance peaks which, however, were absent in our samples.
It remains to be shown whether this is due to a suppression, e.g.~by strong disorder at the barrier or due to an order para\-meter without sign change.

\acknowledgments
{
J.~Tomaschko gratefully acknowledges support by the Evangelisches Studienwerk e.V. Villigst.
V.~Leca acknowledges partial financial support by the Romanian Ministry of Education and Research (Human Resources Reintegration project number 1476/2006) and by CNCSIS -UEFISCSU (project number PNII - IDEI ID\_743/2007).
This work was funded by the Deutsche Forschungsgemeinschaft (project KL 930/11).
}

\bibliography{SLCO_bibliography}
\end{document}